\documentclass[journal]{IEEEtran}
\usepackage{amsmath,amsfonts}
\usepackage{algorithmic}
\usepackage{algorithm}
\usepackage{array}
\usepackage[caption=false,font=normalsize,labelfont=sf,textfont=sf]{subfig}
\usepackage{textcomp}
\usepackage{stfloats}
\usepackage{url}
\usepackage{verbatim}
\usepackage{graphicx}
\graphicspath{{pics/}}
\usepackage{cite}
\usepackage{orcidlink}
\usepackage{booktabs}
\usepackage{placeins}
\usepackage{xcolor}
\begin{document}

\title{Do Ionogram Images Improve Short-Term Forecasts of Post-Sunset Equatorial GNSS Scintillation?}

\author{Mert~Can~Turkmen~\orcidlink{0009-0003-6195-1715},
Septi~Perwitasari~\orcidlink{0000-0002-5108-0077},
Eng~Leong~Tan~\orcidlink{0000-0003-3003-6007},
Pornchai~Supnithi,
Le~Truong~Thanh,
and~Yee~Hui~Lee~\orcidlink{0000-0001-6452-9606}
\thanks{M. C. Turkmen, E. L. Tan, and Y. H. Lee are with the School of
Electrical and Electronic Engineering, Nanyang Technological University
(NTU), Singapore 639798 (e-mail: \mbox{mertcan001@e.ntu.edu.sg};
\mbox{eeltan@ntu.edu.sg}; \mbox{eyhlee@ntu.edu.sg}).}
\thanks{S. Perwitasari is with the National Institute of Information and
Communications Technology (NICT), Koganei, Tokyo, Japan (e-mail:
\mbox{septi.perwitasari@nict.go.jp}).}
\thanks{P. Supnithi is with the School of Engineering, King Mongkut's
Institute of Technology Ladkrabang (KMITL), Bangkok, Thailand (e-mail:
\mbox{pornchai.su@kmitl.ac.th}).}
\thanks{L. T. Thanh is with the Institute of Earth Sciences, Vietnam Academy
of Science and Technology (IES-VAST), Hanoi, Vietnam (e-mail:
\mbox{igpthanh@gmail.com}).}
\thanks{Corresponding author: M. C. Turkmen (e-mail:
\mbox{mertcan001@e.ntu.edu.sg}).}
}

\markboth{Preprint}%
{Turkmen \MakeLowercase{\textit{et al.}}: Do Ionogram Images Improve Short-Term Forecasts of GNSS Scintillation?}

\maketitle

\begin{abstract}
Short-term forecasts of post-sunset equatorial scintillation on Global Navigation Satellite System (GNSS) mostly use GNSS observations and background drivers. Some studies on day-to-day forecasts have used ionosonde data, such as the F-layer virtual height and its time derivatives, as precursors for spread-F and plasma bubbles. Extending these approaches, we explore whether the full ionogram image improves short-term scintillation forecasts related to bubbles beyond GNSS observations. Data come from Chumphon and Bac Lieu, two equatorial Southeast Asian stations, over 2022--2026. Three machine learning models use 2~h of GNSS data and background drivers (Scalar model), ionograms (Image model), or both (Fusion model) to forecast station-level scintillation over the next 2~h. A phase analysis of 230 test nights finds the largest ionogram gain between event onset and peak, where the Image and Fusion models significantly improve the Brier skill score over the Scalar model by 26\% and 23\%. Shapley values show the ionogram raises the event probability in the hour before onset, mainly through the low-frequency end of the F trace during the pre-reversal enhancement period. Adding virtual height and its rate of change to the Scalar model recovers less than half of the ionogram gain. This suggests the image carries richer information, such as early spreading. Finally, we evaluate 2~h lead forecasts with all test events aligned on onset. The Scalar model with virtual-height-derived parameters has no skill until the scintillation index enters its input window, whereas the Image and Fusion models show skill from about 40~min before onset, with 28--36\% lower probability errors.
\end{abstract}


\section{Introduction}
\label{sec:intro}
\IEEEPARstart{I}onospheric delay in Global Navigation Satellite System (GNSS) is removed by dual-frequency combinations or reduced by broadcast correction models for single-frequency users~\cite{Morton20}. However, the ionosphere can still significantly degrade GNSS operations~\cite{Demyanov21,Yang25}. Near the magnetic equator, one cause of this degradation is equatorial plasma bubbles (EPBs), regions of the ionosphere where the plasma is strongly depleted relative to the surrounding medium~\cite{Woodman76,McClure77}. They form after sunset at the bottom of the F layer, rise through it, and stretch along the Earth's magnetic field lines~\cite{Bha22}. Inside them, the plasma may break up into smaller irregularities that scatter radio waves. The resulting rapid fluctuations in signal amplitude and phase are called scintillation~\cite{Kintner07}. Unlike the delay, scintillation cannot be removed by correction models and can only be partially mitigated~\cite{Morton20}. Thus, limiting this impact requires understanding and predicting when and where EPBs occur.

EPBs can be observed with both GNSS receivers and ionosondes. For GNSS, they appear as depletions in total electron content (TEC) and as irregularity signatures quantified by scintillation indices such as $S_4$~\cite{Kintner07,Nishioka08}. On ionograms, the irregularities spread the normally sharp F-layer echo into diffuse traces, a signature known as equatorial spread F (ESF)~\cite{Berkner34,Booker38,Cohen61,Kil15}.

The seasonal and solar-cycle climatology of EPBs is well established~\cite{Stolle24, Andoh26}, but predicting their day-to-day occurrence and onset remains challenging~\cite{Kil15,Li21}. Their formation is commonly attributed to the Rayleigh--Taylor instability~\cite{Sultan96}. The instability is favored by the pre-reversal enhancement (PRE), a brief increase in upward F-layer plasma drift around sunset~\cite{Farley86,Abdu03}. The PRE lifts the F layer to heights where the instability grows faster. Studies have therefore used the ionosonde-derived F-layer virtual height ($h'F$) and its time derivatives as precursors for forecasting day-to-day ESF or EPB occurrence~\cite{Anderson17,Abadi23,Patra25}. Machine learning (ML) has extended this approach to short-term ESF forecasting by combining $h'F$ and other scaled parameters with background drivers~\cite{Thammavongsy23}.

ML has also been used to predict scintillation directly, including day-to-day occurrence and intensity~\cite{Zha21,Mua26}, single-station $S_4$ estimation and short-term prediction~\cite{Tra24,Car22}, and regional $S_4$ map forecasting~\cite{Ata24,Gao25b}. These models use GNSS observables, often combined with background drivers such as solar and geomagnetic indices. As in the ESF and EPB models above, the models that include ionosonde data use scaled parameters such as $h'F$ and $h_mF_2$~\cite{Zha21,Car22}.

Computer vision models such as convolutional neural networks (CNNs) make it practical to use ionograms as whole images. Such models have been applied to ionogram images for spread-F detection and classification~\cite{Wang23,Benchawattananon24} and for ionogram morphology forecasting~\cite{Gao24,Qiu25,Cai26}. Multimodal studies have also combined ionogram images with background drivers~\cite{Sartipzadeh26} or global TEC maps~\cite{Prasad26} to estimate ionospheric parameters. In both studies, ablations showed that adding the ionogram improved the average performance. However, to our knowledge, ionogram images have not been used to directly forecast post-sunset scintillation related to EPBs. It is also unclear when the ionogram contributes to a forecast, which part of the image matters, and whether scaled parameters such as $h'F$ can replace it.

This study explores whether ionograms carry information for short-term scintillation forecasting that GNSS observables do not. Three models are compared: a scalar-only model using GNSS and background drivers, an image-only model using ionogram sequences, and a Fusion model that combines the two. The models share the same decoder, state size, and training procedure, so differences in skill can be attributed to the input modality. The analysis first compares the overall skill of the three models and their skill in each event phase. Then we study which ionogram frames and image regions carry the image contribution, and how the contribution of each input changes as the onset approaches using an explainability method. Finally, we test whether adding $h'F$ and its time derivatives to the scalar-only model can replace the image.

\section{Data}
\label{sec:data}
We use data from the Southeast Asia Low-latitude Ionospheric Network (SEALION), operated by the National Institute of Information and Communications Technology (NICT)~\cite{maruyama2007}. Two SEALION stations are selected, each with a co-located ionosonde and GNSS scintillation receiver: Chumphon (CPN, 10.72$^{\circ}$N, 99.37$^{\circ}$E), Thailand, and Bac Lieu (BCL, 9.30$^{\circ}$N, 105.71$^{\circ}$E), Vietnam. Both stations lie near the magnetic equator, about 700\,km apart. The ionosondes are frequency-modulated continuous-wave (FMCW) sounders that sweep from 2 to 30\,MHz~\cite{nozaki2009}. The GNSS receivers are Septentrio PolaRx5 units. They have operated since 2019, but scintillation was rare at both stations before 2022~\cite{Andoh26}. Both ionosondes were also offline for more than a year, until September 2022 at CPN and October 2022 at BCL. We therefore use data from September 2022 to May 2026.

\subsection{Ionogram}
\label{sec:ionogram}

\begin{figure}[!htbp]
\centering
\includegraphics[width=\columnwidth]{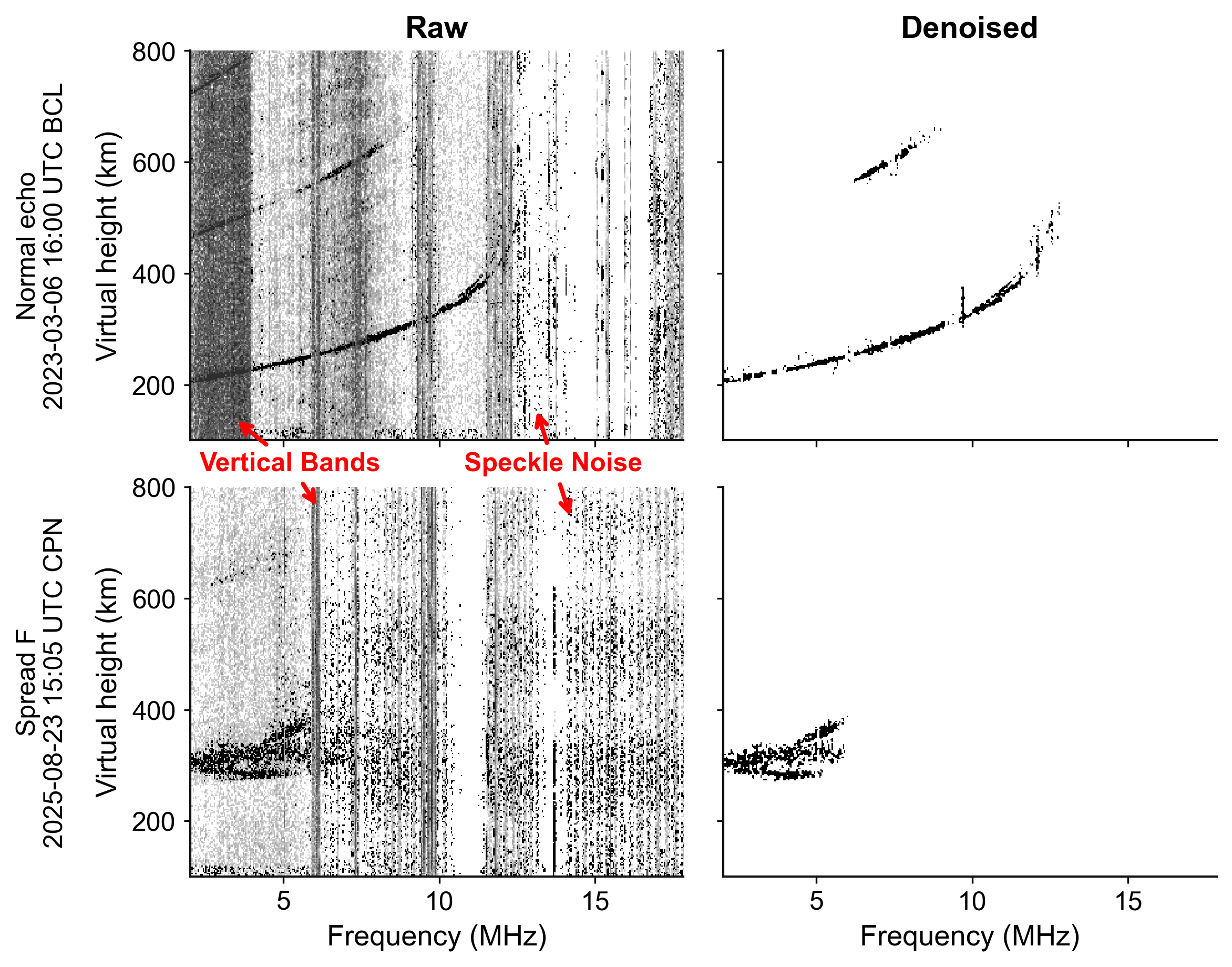}
\caption{Example of denoising of ionograms for a normal echo trace (top, BCL, 2023-06-07 11:15~UTC) and a spread-F case (bottom, CPN, 2025-08-23 15:05~UTC). Red arrows depict the noise types encountered in the raw images.}
\label{fig:ionogram_noise}
\end{figure}

NICT archives one sounding from each ionosonde every 5\,min. Images are cropped to frequencies of 2--17\,MHz and virtual heights of 100--800\,km to focus on the F region. The raw images contain vertical bands from radio interference~\cite{nozaki2009} and speckle noise. The speckle noise can resemble the diffuse spread-F echo and has the same intensity as the echo trace, making separating them difficult with existing methods. Some frames contain only noise and no echo. The model must learn from the echo rather than the noise, so that we can tell whether it uses the information the echo carries. We therefore developed our denoising algorithm. The vertical bands are removed by subtracting the per-column median, and the speckle noise is reduced with a $2 \times 2$ median filter. The pixels that remain are grouped into clusters with Density-Based Spatial Clustering of Applications with Noise (DBSCAN)~\cite{Ester96}. A cluster is then discarded if it is small, loosely packed, or forms a narrow vertical stripe. The clusters that are kept form the denoised binary ionogram, which is resized to $128 \times 128$ pixels as model input. Two examples are shown in Fig.~\ref{fig:ionogram_noise} for a normal echo and spread case. A frame left without an echo trace after denoising is treated as missing. The full algorithm is given in \url{https://github.com/Mert-chan/ionogram-denoiser} and further described in~\cite{tur26b}.

\subsection{GNSS and Background Drivers}
\label{sec:gnss}

\begin{figure}[!htbp]
\centering
\includegraphics[width=\columnwidth]{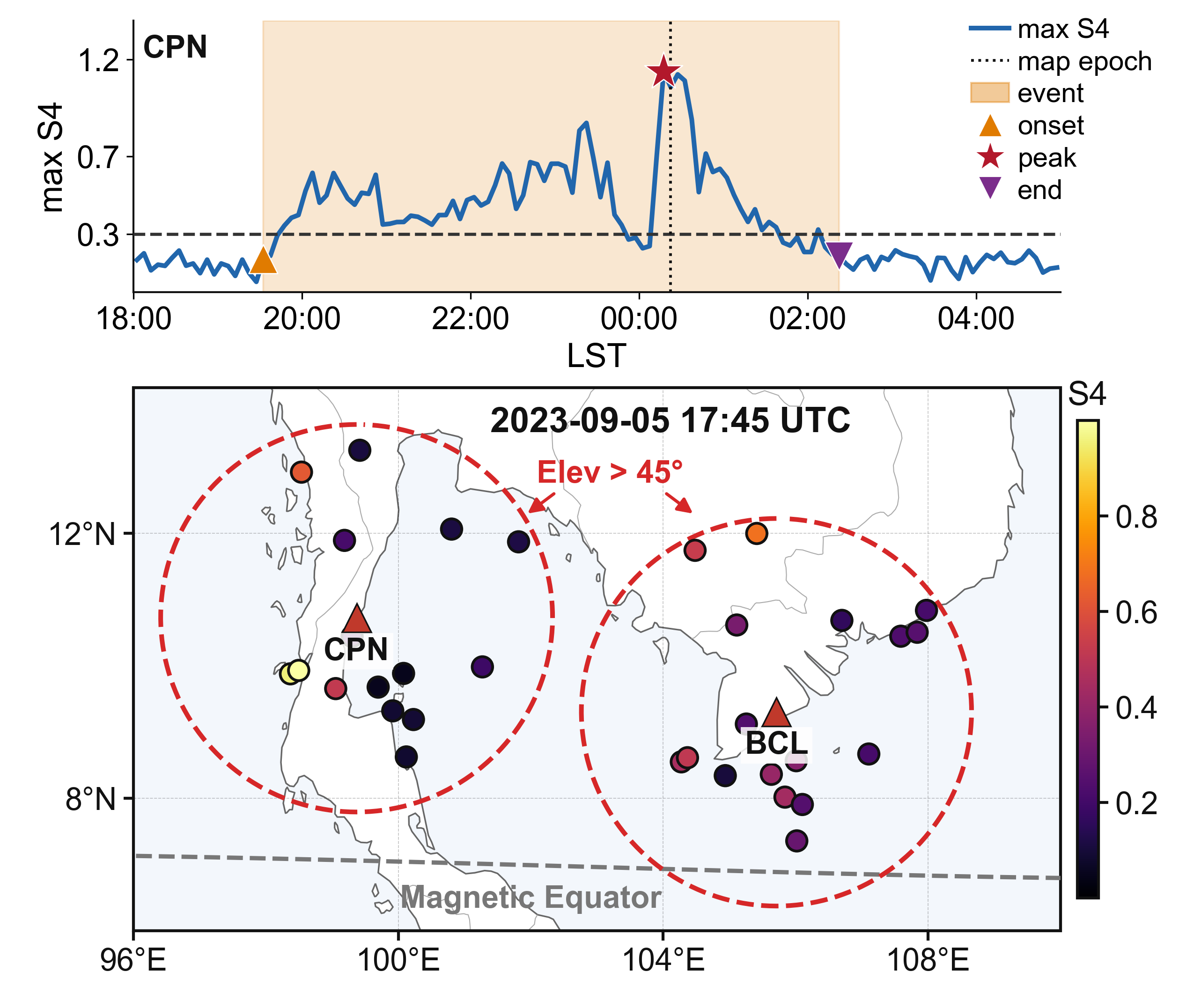}
\caption{Top: time series of the station $S_4$ at CPN on 2023-09-05, with the detected event intervals. Bottom: IPPs above $45^{\circ}$ elevation around CPN and BCL, colored by $S_4$ (17:45~UTC on the same night).}
\label{fig:station_map}
\end{figure}

Each receiver records the $S_4$ index and vertical TEC (VTEC) per line of sight (LoS) at 1\,min resolution for GPS, Galileo, BeiDou, QZSS, and GLONASS.
Satellite positions come from the Wuhan University multi-GNSS precise orbit products~\cite{Li22}. The ionospheric pierce point (IPP) of each LoS is computed on a thin shell at 350\,km altitude. Only LoS above $45^{\circ}$ elevation are kept, limiting the pierce points to about 320\,km from the station (Fig.~\ref{fig:station_map}, bottom). 
The LoS data are resampled to the 5\,min ionogram cadence. We then take the maximum $S_4$ over all LoS to derive the event labels. This gives one $S_4$ time series per station.

Scintillation events are detected similarly to~\cite{tur26} but for the period of 16:00 to 06:00 local solar time (LST) on the station $S_4$. A run of samples above the night baseline is an event if at least three consecutive samples exceed the event threshold of $S_4 = 0.3$, the lower bound of weak scintillation in~\cite{Aguiar25}. We then trace backward and forward from the peak to the first and last baseline samples, which mark the onset and the end of the event (Fig.~\ref{fig:station_map}, top). All time steps between the onset and the end are labeled as part of the event. The models forecast this event label.

For the background drivers, the hourly disturbance storm-time index (Dst) and the daily 10.7\,cm solar radio flux (F10.7) are taken from the NASA OMNIWeb database~\cite{King05} and assigned to each 5\,min sample. The LST is also computed at the station coordinates and encoded cyclically as $\sin(2\pi t/24)$ and $\cos(2\pi t/24)$, with $t$ the LST in hours.

\FloatBarrier

\subsection{Dataset}
\label{sec:dataset}

\begin{figure*}[!htbp]
\centering
\includegraphics[width=\textwidth]{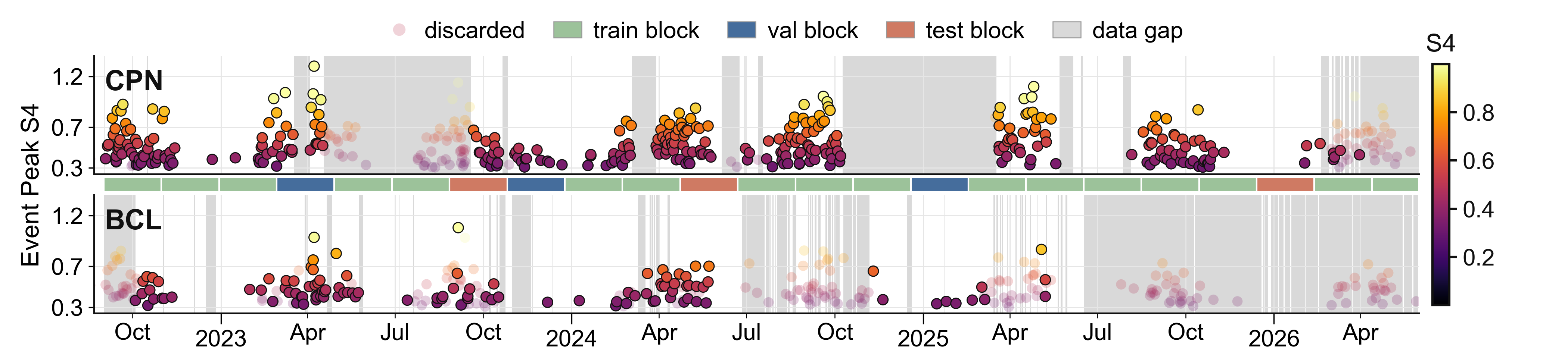}
\caption{Scintillation events at CPN and BCL from September 2022 to May 2026, shown by event peak $S_4$. The bar between the panels shows the 60-night training, validation, and test blocks, which are common to both stations. Gray marks data gaps at each station. Faded markers are events dropped because of missing data.}
\label{fig:events_timeline}
\end{figure*}
Fig.~\ref{fig:events_timeline} shows the detected events and the data gaps. Gaps are frequent and most of them come from the ionosonde. GNSS data are available for about 86\% of the nighttime time steps, ionograms for about 65\%, and both together for 58\%. The ionogram loss has two causes: instrument outages, over both short and extended periods, and frames that hold no echo trace after denoising (Section~\ref{sec:ionogram}).

Each sample is a 4\,h overlapping window within one night (2\,h input, 2\,h prediction, 5\,min steps). Only windows with both data types are used, so that all three models are trained and tested on the same samples. A window is kept if, for each data type, at most 20\,min is missing. Missing ionograms are replaced by an empty image and flagged by the empty-image mask. Events whose 2\,h input window misses more than 20\,min of images, or whose night has no kept window, are dropped (faded markers in Fig.~\ref{fig:events_timeline}). This leaves 436 of the 731 detected events, on 409 station-nights.

Because EPB occurrence is strongly seasonal and the data gaps are uneven, a chronological split would be heavily unbalanced. We instead cut the record into blocks of 60 consecutive nights, common to both stations, and assign them to training, validation, and test (Fig.~\ref{fig:events_timeline}). The assignment focuses on the events and divides them close to 70:15:15, with a similar spread of $S_4$ in each subset. Samples never cross the daytime gap (06:00--16:00 LST), so subsets are separated by at least 10\,h. Table~\ref{tab:dataset_split} lists the resulting data split statistics, with an event rate of 10.7--12.7\% in each subset.

\begin{table}[t]
\centering
\caption{Dataset Statistics.}
\label{tab:dataset_split}
\begin{tabular}{lcccc}
\toprule
 & Train & Validation & Test & Total \\
\midrule
Station-nights & 1,041 & 207 & 230 & 1,478 \\
Events & 306 & 66 & 64 & 436 \\
Samples & 100,872 & 17,741 & 21,972 & 140,585 \\
Event rate (\%) & 12.7 & 10.8 & 10.7 & 12.1 \\
\bottomrule
\end{tabular}

\smallskip
{\footnotesize Counts are over both stations. The event rate is the share of prediction time steps labeled as event.}
\end{table}

\section{Method}
\label{sec:method}

\begin{figure*}[!htb]
\centering
\includegraphics[width=\textwidth]{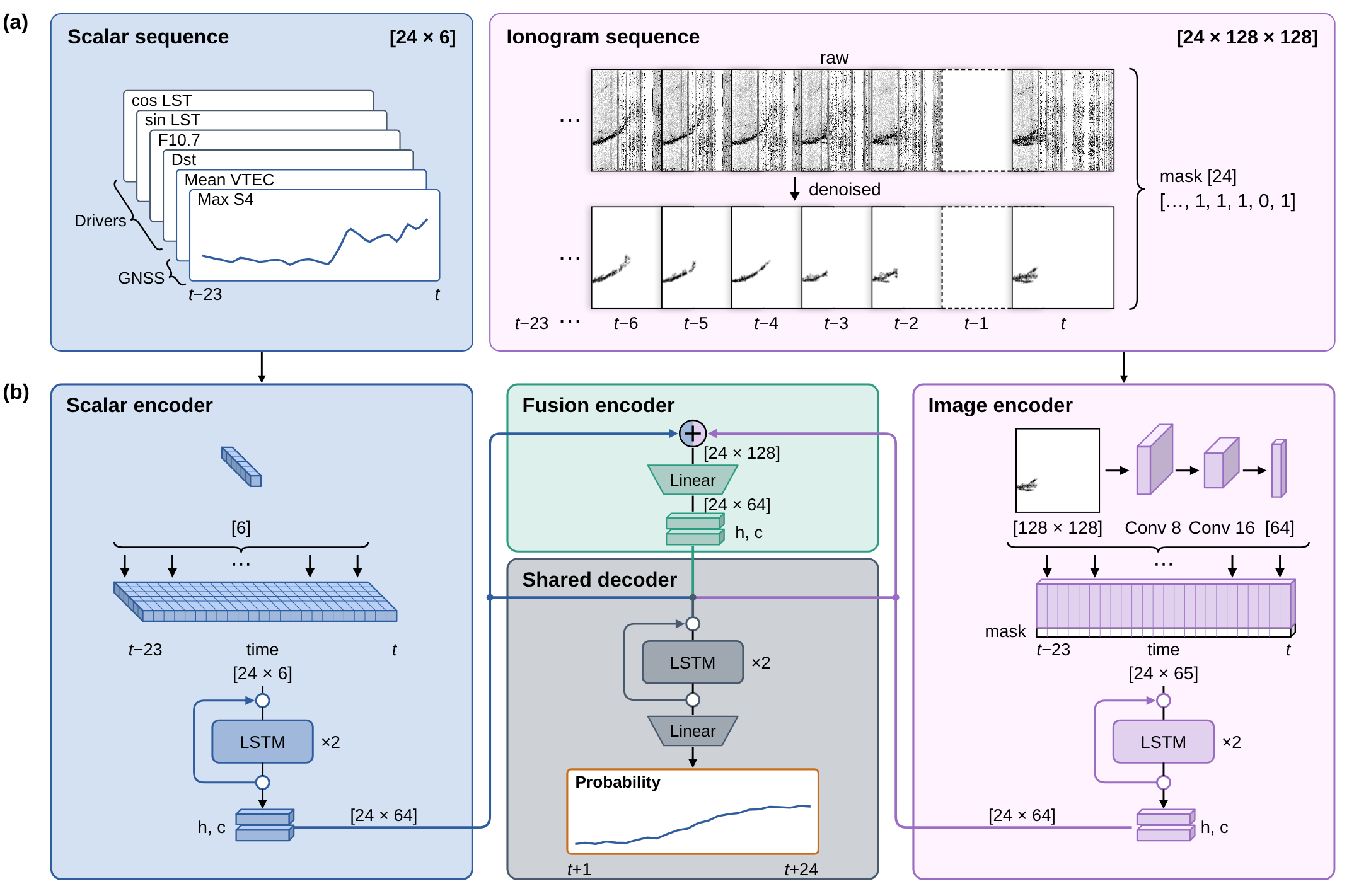}
\caption{(a) Scalar and ionogram input sequences. (b) The model encoders and their shared decoder. Both encoders are two-layer LSTMs, and the ionogram encoder first embeds each frame with a CNN. The Fusion encoder merges the final states of both encoders with a linear layer.}
\label{fig:models}
\end{figure*}

\subsection{Models}
\label{sec:models}
The models are kept simple, because the aim is to measure what ionogram images add, not to reach the best possible forecast. All three models share the same decoder and differ only in their encoder. The encoder summarizes the 2\,h input window into a hidden and cell state of size 64. A long short-term memory (LSTM) decoder turns this state into the event probability for each of the next 24 steps.

Each model is named after the input its encoder takes (Fig.~\ref{fig:models}, panel (b)). The Scalar model encodes the six scalar inputs with a two-layer LSTM. These inputs are the maximum $S_4$, the mean VTEC, Dst, F10.7, and the sine and cosine of the LST (Fig.~\ref{fig:models}, panel (a)). The Image model embeds each denoised ionogram with a small CNN, appends the empty-image mask, and passes the sequence to a two-layer LSTM of the same size. The Fusion model runs both encoders and merges their final states with a single linear layer. The scalar inputs include the current station $S_4$, so the Fusion model must improve on a baseline that already knows the recent scintillation level.

\subsection{Training and Testing}
\label{sec:training}
All models are trained with the binary cross-entropy loss over the 24 lead times, using the Adam optimizer with a fixed batch size of 64. For this batch size, we pick the learning rate of each model with the learning-rate range test. Training is stopped early when the validation loss does not improve for 10 epochs.

The models output probabilities, and an operational user would turn them into alarms with a threshold that depends on the application. The comparison therefore uses threshold-independent metrics, computed on the test set for each lead time. The main metric is the Brier Skill Score (BSS).

The Brier Score is the mean squared difference between the forecast probability $p_j$ and the observed outcome $y_j \in \{0,1\}$ over all $n$ scored samples,
\begin{equation}
\mathrm{BS} = \frac{1}{n}\sum_{j=1}^{n}(p_j - y_j)^2, \qquad \mathrm{BSS} = 1 - \frac{\mathrm{BS}}{\mathrm{BS}_{\mathrm{clim}}},
\end{equation}
where $\mathrm{BS}_{\mathrm{clim}}$ is the Brier Score of a climatological forecast that always predicts the event rate of the training set. The BSS is the fraction by which a model lowers the Brier Score of this constant forecast. A BSS of 0.5 means that the squared error is half that of climatology, and a BSS of 1 is a perfect forecast. A BSS of 0 means no gain over climatology. A negative BSS means that the model does worse than always forecasting the training event rate of 12.7\%. Since the BSS uses the probabilities directly, it penalizes both poor discrimination and poor calibration. For two models scored on the same samples, it shows which model has the smaller probability error.

We also report two ranking metrics. The area under the receiver operating characteristic curve (ROC-AUC) and the area under the precision--recall curve (PR-AUC).

Along with the metrics we quantify the uncertainty of each score with a nightly bootstrap. The test nights are resampled with replacement 1,000 times, and every metric is recomputed on each resample. Reported values are the mean and standard deviation over the resamples. Where relevant, we also use a bootstrap using only event nights. Two models are compared with a paired bootstrap, in which both are scored on the same resampled nights or events, and a gain is significant when $p < 0.05$.

\subsection{Explainability}
\label{sec:explainability}
This study uses Shapley values~\cite{Lundberg17} to identify when and where the ionogram contributes to a forecast. To calculate, the method treats parts of the input as players in a cooperative game and splits a single forecast into contributions from the set $F$ of players,
\begin{equation}
P(x) = \phi_0 + \sum_{i \in F} \Phi_i,
\label{eq:shapley}
\end{equation}
where $P(x)$ is the predicted event probability, $\phi_0$ is a reference probability, and $\Phi_i$ is the contribution of player $i$. The contribution of a player is the change in the forecast when the player is added, averaged over all orders in which the players can be added. A player that is left out takes its values from reference windows. For the reference, we draw 16 quiet test nights at random, at the same station and local time. Then $\phi_0$ becomes the mean forecast for a quiet night at that station and time, and $\Phi_i$ is how much player $i$ raises or lowers the event probability relative to it.

We define the players in two levels. At the group level, the players are the model inputs, split into the ionogram and three scalar groups,
\begin{equation}
\begin{aligned}
F &= \{\mathrm{ionogram}\} \cup F_{\mathrm{s}}, \\
F_{\mathrm{s}} &= \{S_{4,\mathrm{max}},\, \mathrm{VTEC}_{\mathrm{mean}},\, \mathrm{drivers}\},
\end{aligned}
\end{equation}
where the 2-h input window of ionograms counts as one player. The 2-h input sequence of the drivers Dst, F10.7, and LST also counts as one player. A capital $\Phi_i$ therefore denotes the contribution of a whole 2-h input sequence. Four players give $4! = 24$ orderings, all of which are evaluated.

The contribution of a single frame, and of a region within it, needs a different design. We split each $128 \times 128$-pixel frame into blocks of $4 \times 4$ pixels (a $32 \times 32$ grid), about 0.5~MHz $\times$ 22~km each. Block $k$ of frame $t$ is a player with contribution $\phi_{t,k}$. A block that is left out takes the same block of the same frame from each reference window. The scalar groups in $F_{\mathrm{s}}$ remain players, so equation~\eqref{eq:shapley} becomes
\begin{equation}
P(x) = \phi_0 + \sum_{t=1}^{24} \sum_{k=1}^{1024} \phi_{t,k} + \sum_{i \in F_{\mathrm{s}}} \Phi_i.
\end{equation}
The $\phi_{t,k}$ of one frame form a map of the ionogram (Fig.~\ref{fig:shap_frames}, left), and their sum $\phi_t = \sum_k \phi_{t,k}$ is the contribution of the whole frame (Fig.~\ref{fig:shap_frames}, right). With 24,579 players, the contributions cannot be computed exactly; instead, they are estimated by sampling over 50 random player orderings~\cite{Castro09}. An independent run with different random orderings reproduces $\phi_{t,k}$ and $\phi_t$ with correlations of $r = 0.975$ and $0.998$, respectively. Individual blocks within the frame carry slightly more sampling noise than the whole-frame totals, but this does not affect our main results in Section~\ref{sec:when}.

\section{Results}
\label{sec:results}

\subsection{Overall Results}

\begin{table}[t]
\centering
\caption{Overall Results on the Test Set}
\label{tab:main_results}
\begin{tabular}{@{}lccc@{}}
\toprule
Model & BSS & PR-AUC & ROC-AUC \\
\midrule
Image & $0.442 \pm 0.045$ & $0.666 \pm 0.052$ & $0.942 \pm 0.010$ \\
Scalar   & $0.474 \pm 0.034$ & $0.730 \pm 0.039$ & $0.934 \pm 0.012$ \\
Fusion   & $\mathbf{0.499 \pm 0.041}$ & $\mathbf{0.750 \pm 0.041}$ & $\mathbf{0.952 \pm 0.008}$ \\
\bottomrule
\end{tabular}

\smallskip
{\footnotesize Values are mean $\pm$ standard deviation. Best value of each metric is in bold.}
\end{table}

Table~\ref{tab:main_results} shows the test-set skill of the three models on 230 test nights from both stations, with all 24 lead times (5--120~min) pooled and scores averaged over three training seeds. Although the Image model is the weakest model in BSS and PR-AUC, it reaches a PR-AUC of 0.67 and a ROC-AUC of 0.94 from images alone. The Scalar model, with all the GNSS observables and standard drivers, is stronger than the Image model in BSS (0.47) and PR-AUC (0.73), though slightly weaker in ROC-AUC (0.93). The Fusion model is the strongest on all three metrics, with a BSS of 0.50, a PR-AUC of 0.75 and a ROC-AUC of 0.95, but its gains over the Scalar model are small. Most time steps are quiet, so they dominate the pooled scores and can hide differences during the events. Therefore we next examine the skill in the phases of the events.

\subsection{Event-Phase Results}

\begin{table}[t]
\centering
\caption{Event-Phase Results}
\label{tab:phase_results}
\begin{tabular}{llcc}
\toprule
Model & Metric & Onset to peak & Peak to end \\
\midrule
         & BSS     & $\mathbf{0.522 \pm 0.047^{*}}$ & $0.506 \pm 0.037$ \\
Image    & PR-AUC  & $0.893 \pm 0.029$ & $0.824 \pm 0.029$ \\
         & ROC-AUC & $0.750 \pm 0.039$ & $0.740 \pm 0.027$ \\
\midrule
         & BSS     & $0.413 \pm 0.054$ & $0.518 \pm 0.042$ \\
Scalar   & PR-AUC  & $0.898 \pm 0.030$ & $0.867 \pm 0.025$ \\
         & ROC-AUC & $0.742 \pm 0.039$ & $0.774 \pm 0.029$ \\
\midrule
         & BSS     & $0.509 \pm 0.051^{*}$ & $\mathbf{0.542 \pm 0.044}$ \\
Fusion   & PR-AUC  & $\mathbf{0.901 \pm 0.029}$ & $\mathbf{0.869 \pm 0.024}$ \\
         & ROC-AUC & $\mathbf{0.763 \pm 0.038}$ & $\mathbf{0.783 \pm 0.027}$ \\
\bottomrule
\end{tabular}

\smallskip
{\footnotesize Values are mean $\pm$ standard deviation of night bootstrap. Best value of each metric is in bold. $^{*}p < 0.05$ against the Scalar model.}
\end{table}

\begin{figure}[!htb]
\centering
\includegraphics[width=\columnwidth]{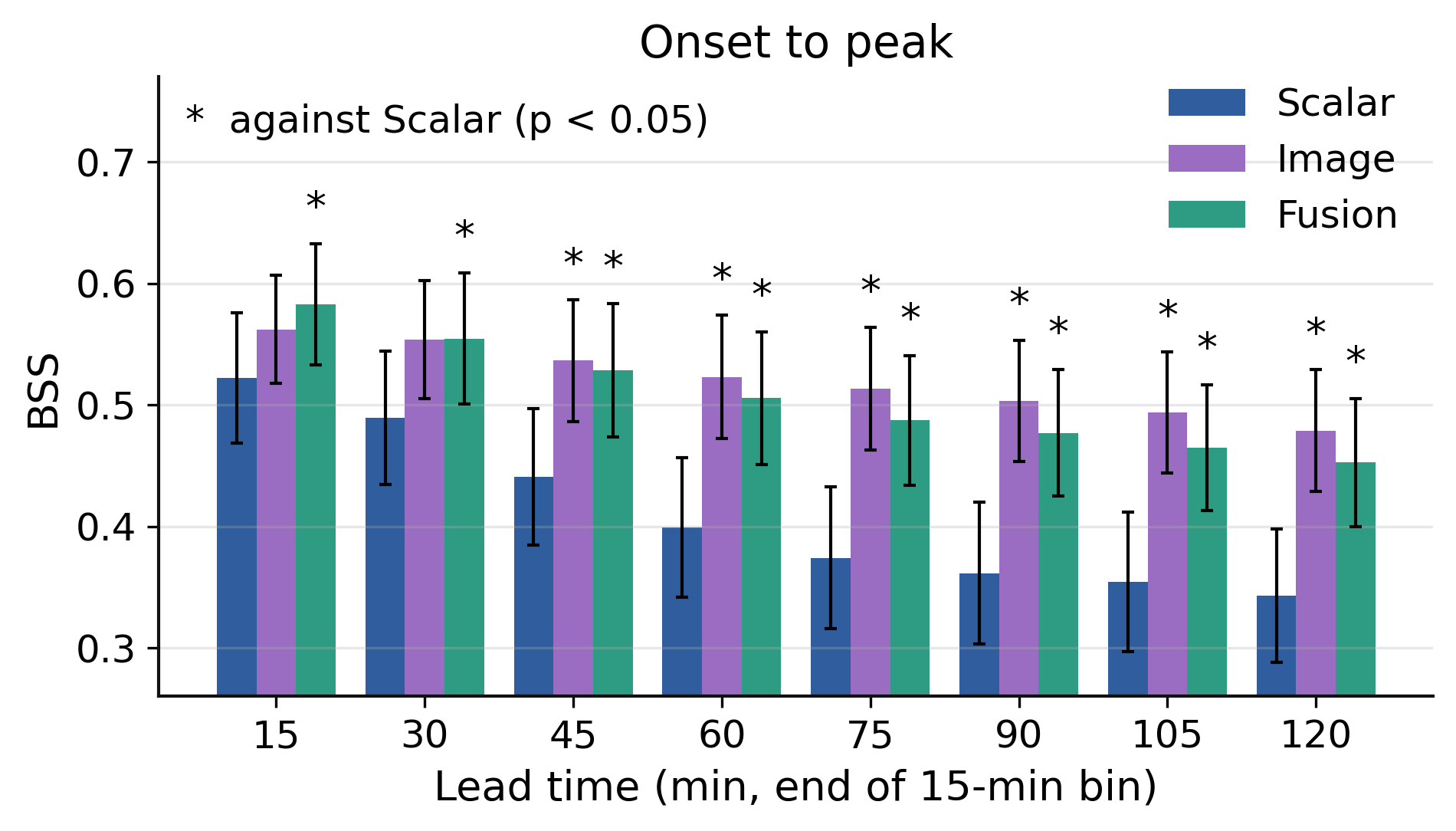}
\caption{BSS of the three models in the onset to peak phase per lead time. Lead times are grouped into 15-min bins and labeled by the end of the bin. Error bars show the standard deviation, and asterisks mark a significant gain over the Scalar model.}
\label{fig:phase_lead}
\end{figure}

We evaluate forecast skill in two phases defined as: 1~h before onset to the peak (onset to peak) and the peak to 1~h after the event ends (peak to end), where the 1~h buffers include quiet samples in each phase. Table~\ref{tab:phase_results} summarizes the pooled performance across all lead times.

The ionogram's added value is concentrated before the peak. From onset to peak, the BSS increases significantly from 0.41 (Scalar) to 0.52 (Image, $+26\%$) and 0.51 (Fusion, $+23\%$). PR-AUC and ROC-AUC show no significant differences across models in either phase, indicating that the ionogram gain comes from better-calibrated event probabilities rather than ranking. From peak to end, all three models perform similarly (0.52 Scalar, 0.51 Image, and 0.54 Fusion).

Resolving the onset to peak phase by lead time shows that the ionogram advantage widens substantially at longer horizons (Fig.~\ref{fig:phase_lead}). The Scalar model loses over a third of its skill between 15 and 120~min, falling from a BSS of 0.52 to 0.34 ($-34\%$). By contrast, the Image model declines much more slowly, from 0.56 to 0.48 ($-15\%$); at 105--120~min (0.49--0.48), it still matches the Scalar model's skill at 30~min (0.49). At short leads (15--30~min), recent GNSS observations remain informative, so only the Fusion model significantly outperforms Scalar (by 12--13\%). From 45~min onward, both ionogram models significantly beat Scalar, with relative gains of 31--40\% for Image and 27--32\% for Fusion across 60--120~min. At the 120-min lead, Image (0.48) and Fusion (0.45) exceed Scalar (0.34) by 40\% and 32\%, respectively.

\subsection{When and Where the Ionogram Contributes}
\label{sec:when}

\begin{figure*}[!htb]
\centering
\includegraphics[width=\textwidth]{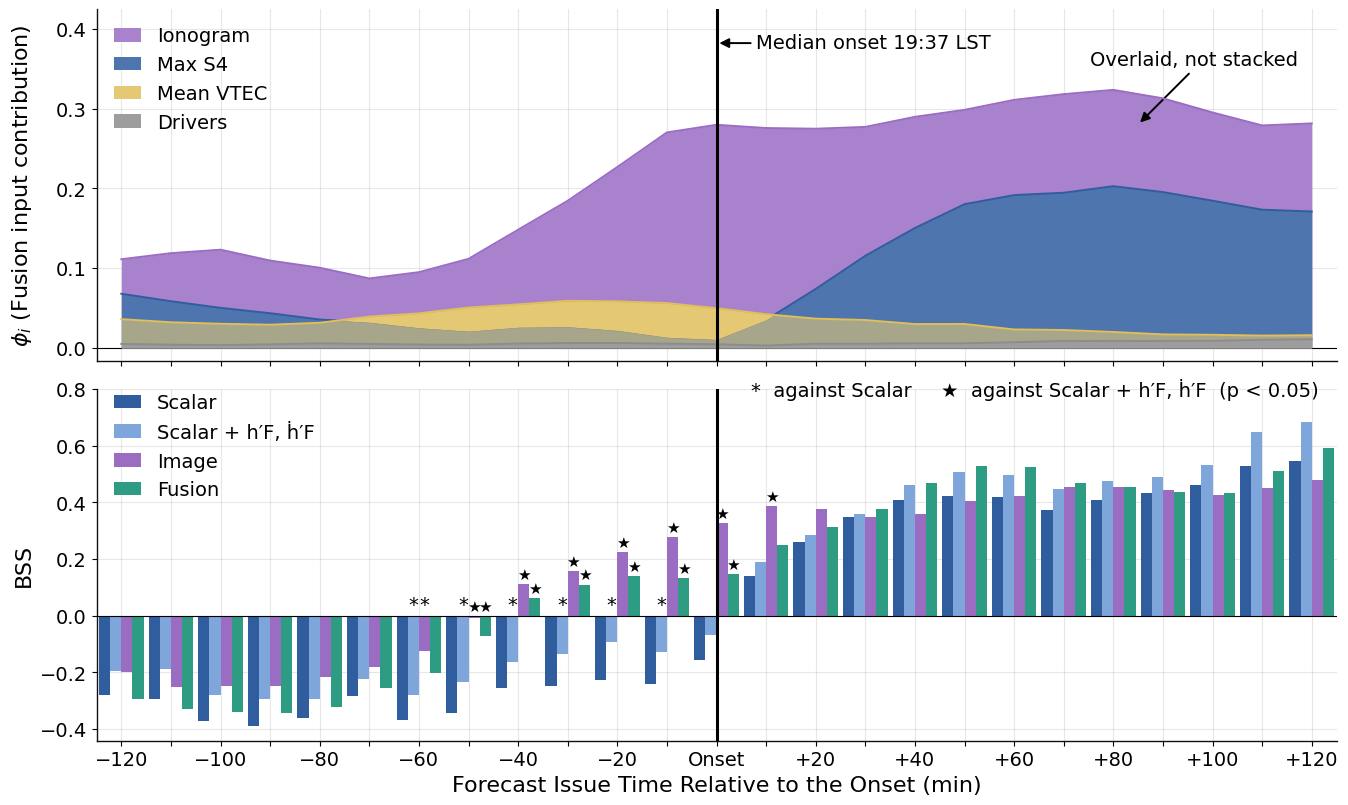}
\caption{ Top: Shapley contributions $\Phi_i$ of the four input groups to the Fusion forecast probability, averaged over all lead times and test events. Bottom: BSS of the four models for the 120~min lead time. Asterisks mark a significant gain over the Scalar model, and stars a significant gain over Scalar + $h'F$, $\dot{h}'F$. Both panels use 10-min bins that each average two 5-min issue times.}
\label{fig:when}
\end{figure*}

\begin{figure*}[!htb]
\centering
\includegraphics[width=\textwidth]{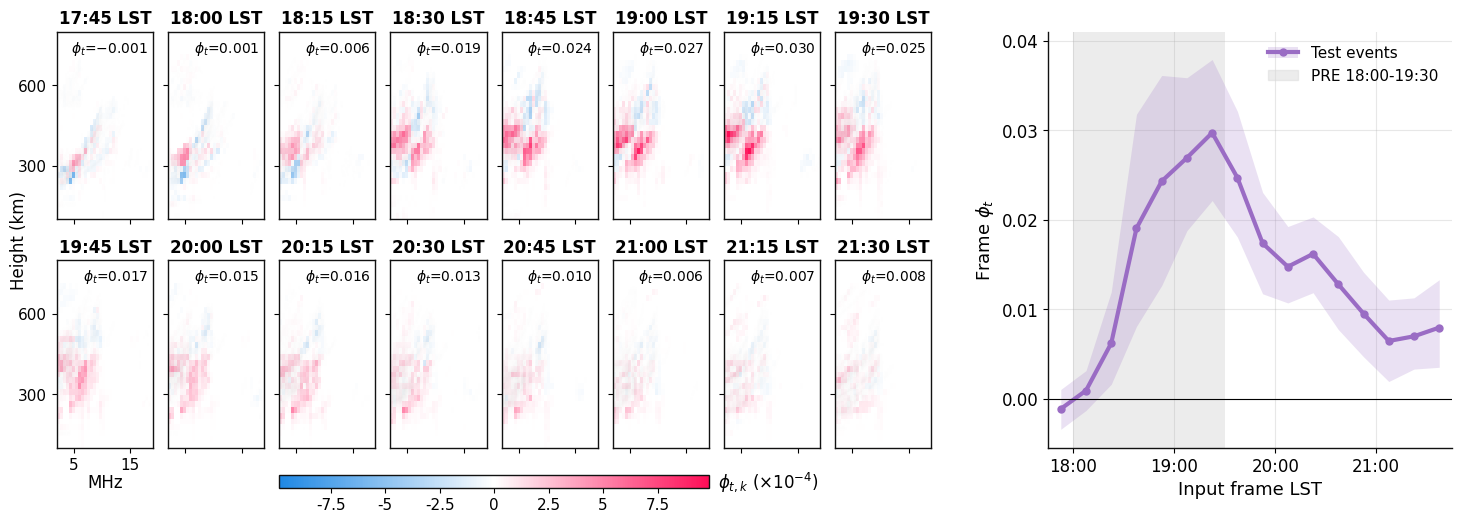}
\caption{Block-level Shapley contributions of the ionogram frames to the Fusion forecast at the 60-min lead, averaged over the test events. Left: mean block contribution $\phi_{t,k}$ on the frequency--height grid, for input frames in 15-min LST bins from 17:45 to 21:30~LST. Red blocks raise the event probability and blue blocks lower it. The number in each map is the frame contribution $\phi_t$. Right: $\phi_t$ against the LST of the input frame where the gray band marks the PRE window (18:00--19:30~LST).}
\label{fig:shap_frames}
\end{figure*}

The phase results place the ionogram gain in the period from 1~h before the onset to the peak for test events. They do not show when in this phase the ionogram contributes, or where in the image sequence the contribution comes from. The two Shapley games designed in Section~\ref{sec:explainability} are used to examine both.

First we align every test event on its onset and compute the group-level contributions $\Phi_i$ of the four input groups to the Fusion forecast probability, averaged over all lead times. Fig.~\ref{fig:when} (top) shows these contributions against the forecast issue time relative to the event onset. The ionogram contribution rises from about 0.1 at 60~min before onset to about 0.28 at the onset. Before the onset, the GNSS observables and drivers each contribute less than 0.1. The ionogram contribution is therefore concentrated in the hour before the onset. For the median onset at 19:37~LST, this hour corresponds to forecasts issued between about 18:30 and 19:30~LST. We further look at the block-level Shapley values to see where the contribution comes from.

Fig.~\ref{fig:shap_frames} (left) shows the block contributions at the 60-min lead, averaged over the test events, for input frames from 17:45 to 21:30~LST. As the LST moves from 18:00 to 19:30, the red blocks, which raise the event probability, become more numerous and darker. This pattern is most apparent between 2 and 10~MHz and between 300 and 450~km. In the frames from 17:45 to 18:15~LST, however, the lower part of the echo trace, where the trace starts, is blue and lowers the probability. It turns red in later frames. This region may be where the model infers the virtual height of the F layer.

The frame contribution $\phi_t$ rises gradually from slightly negative in the 17:45~LST bin to a peak of 0.030 in the 19:15~LST bin (Fig.~\ref{fig:shap_frames}, right). On average, a frame recorded in the 19:15~LST bin raises the event probability by about 3 percentage points relative to the quiet-night reference. This rise coincides with the PRE window (18:00--19:30~LST), which Abadi et al.~\cite{Abadi23} used to forecast ESF in the similar region including both CPN and BCL. After the peak, $\phi_t$ falls to 0.015 at 20:00~LST and stays below 0.01 from 20:45~LST on.

After the PRE window, single frames contribute less (Fig.~\ref{fig:shap_frames}, right), but the contribution of the whole ionogram window stays high after the onset (Fig.~\ref{fig:when}, top). The two are consistent, because the window holds 24 frames and their smaller contributions add up. After the onset, the $S_4$ in the input window also shows that scintillation has started. For the Fusion model, the information the ionogram carries by itself is therefore concentrated in the frames recorded in the PRE window, before the onset, and in the low-frequency part of the echo trace corresponding to the rising F layer.

\subsection{Can the Onset Gain Be Summarized by \texorpdfstring{$h'F$}{h'F}?}
\label{sec:hf}

\begin{figure}[!htb]
\centering
\includegraphics[width=\columnwidth]{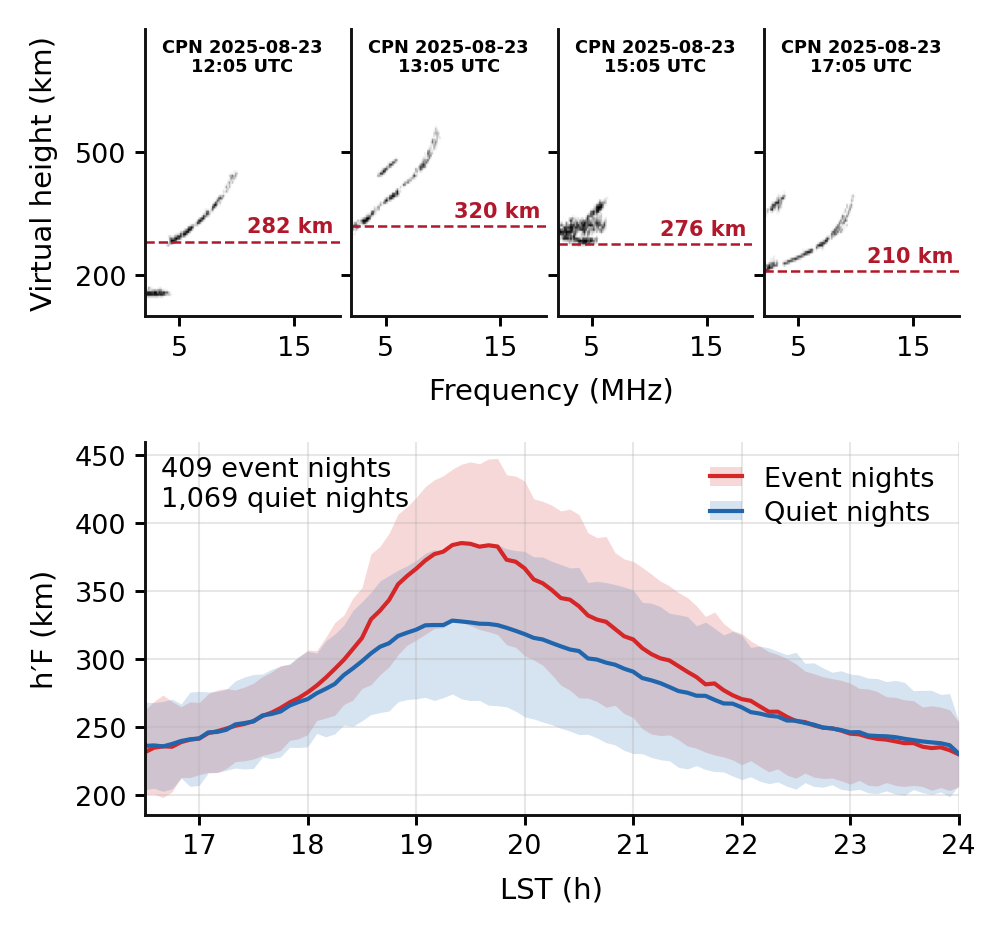}
\caption{Top: denoised CPN ionograms on 2025-08-23 at 12:05, 13:05, 15:05 and 17:05~UTC, with the approximated $h'F$ marked by the dashed line. Bottom: mean $h'F$ against LST for nights with an event and quiet nights, with the shading showing $\pm 1$ standard deviation.}
\label{fig:hf}
\end{figure}

Since the image gains are mostly concentrated in the PRE window, we ask whether the virtual height $h'F$ and its derived parameters can summarize the information in the ionogram. NICT provides scaled $h'F$ every 15~min, which we treat as ground truth. To match the 5~min model cadence, we estimate our own $h'F$ from each denoised ionogram as the lowest echo pixels (Fig.~\ref{fig:hf}, top). We keep the NICT values as reference and add only our change since the last scaled value to fill the 5~min gaps. The resulting $h'F$ series when plotted against LST for both event and quiet nights shows the expected post-sunset rise of the F layer (Fig.~\ref{fig:hf}, bottom). The mean $h'F$ overlaps before about 18:00~LST. After that it rises faster on event nights and peaks near 385~km around 19:25~LST, against about 330~km on quiet nights. The PRE window (18:00--19:30~LST) therefore coincides with the rising $h'F$ on event nights.

The first time derivative, $\dot{h}'F$, is the uplift rate of the layer and approximates the upward plasma drift~\cite{Abadi23}. Patra and Das~\cite{Patra25} used the second, $\ddot{h}'F$, as a marker of local upwelling of the bottomside. Both derivatives are computed following Patra and Das method of smoothing with a 7-point moving average to make sure noise in $h'F$ is minimized. Missing frames stay as gaps in the $h'F$ series and are flagged by the same mask as in the ionogram models. We add these parameters to the inputs of the Scalar model in different combinations (Table~\ref{tab:hf_variants}).

\begin{table}[t]
\centering
\caption{Scalar Model with $h'F$ Inputs}
\label{tab:hf_variants}
\begin{tabular}{lc}
\toprule
Inputs added to the Scalar model & Onset-to-peak BSS \\
\midrule
None                                  & $0.413 \pm 0.054$ \\
\midrule
$h'F$                                 & $0.444 \pm 0.052$ \\
$\dot{h}'F$                           & $0.445 \pm 0.054^{*}$ \\
$\ddot{h}'F$                          & $0.436 \pm 0.055^{*}$ \\
\midrule
$h'F$, $\dot{h}'F$                    & $\mathbf{0.455 \pm 0.052^{*}}$ \\
$h'F$, $\ddot{h}'F$                   & $0.452 \pm 0.052^{*}$ \\
$\dot{h}'F$, $\ddot{h}'F$             & $0.413 \pm 0.053$ \\
\midrule
$h'F$, $\dot{h}'F$, $\ddot{h}'F$      & $0.453 \pm 0.053^{*}$ \\
\bottomrule
\end{tabular}
\vspace{2pt}
\begin{minipage}{\columnwidth}
\smallskip
\footnotesize Values are mean $\pm$ standard deviation from a test event bootstrap. $^{*}p < 0.05$ against the Scalar model.
\end{minipage}
\end{table}

$h'F$ alone raises the onset to peak BSS from 0.413 to 0.444, but the gain is not significant. Adding $\dot{h}'F$ raises it to 0.455, a significant gain over the Scalar model. With all three parameters the BSS is 0.453, and the two derivatives without $h'F$ give 0.413, no gain over the Scalar model. $\ddot{h}'F$ adds nothing on top of $h'F$ and $\dot{h}'F$ (0.453 versus 0.455). The derivatives help only together with $h'F$, and alone they do not improve the forecast. Combining $h'F$ with $\dot{h}'F$ gives the highest BSS of all variants.

The best variant (Scalar + $h'F$, $\dot{h}'F$) still stays below the ionogram models. Its onset to peak BSS of 0.455 compares with 0.522 for the Image model and 0.509 for the Fusion model (Table~\ref{tab:phase_results}). $h'F$ and $\dot{h}'F$ therefore recover less than half of the gain of either ionogram model over the Scalar model. The pooled score does not show where around the onset the remaining gap lies. Fig.~\ref{fig:when} resolves it for the longest lead time of 120~min, where the models must forecast scintillation while the ionosphere is still quiet.

At the 120-min lead, no model has skill for forecasts issued earlier than about 40~min before the onset. The ionogram models gain skill once the frames from this last 40~min enter their input, with a BSS of about 0.2 for the Image model and about 0.1 for the Fusion model. The Scalar model has negative skill compared to climatology until the event has started and the rising $S_4$ is in its input. Adding $h'F$ and $\dot{h}'F$ reduces its negative skill from about 70 to 10~min before the onset. This variant still reaches a positive BSS only after the onset, and it stays significantly below the ionogram models from about 40~min before the onset up to the onset. This implies that the virtual height parameters and GNSS observables do not have forecast skill until the scintillation event starts. For next 2 h forecasts issued in this 40-min window, the Image model and the Fusion model lower the average Brier score over all test events by 36\% and 28\% relative to the Scalar model, and by 29\% and 19\% relative to the variant with $h'F$ and $\dot{h}'F$.

\section{Discussion}
\label{sec:discussion}

The ionogram sees the development of an EPB earlier than the GNSS observables do. Part of this lead comes from the order of events, since a bubble must form before its irregularities can cause scintillation. The ionogram records the evening rise of the F layer that precedes the bubble. The GNSS observables show the event only once irregularities cross the signal paths.

How much of this early signal a model can use depends first on the quality and availability of the ionograms. Both data types are available for only 58\% of the nighttime time steps, and 295 of the 731 detected events were dropped because of missing data (Section~\ref{sec:dataset}). The noise inside the images is another limiting factor. Our denoising method is a heuristic. Where it fails is discussed in~\cite{tur26b}. Therefore, the gain reported here can be a lower bound of what clean ionogram sequences can add. Operational use is not examined in this study, but in a prediction network both the quality of the ionograms and their availability should be considered in the forecast model.

Comparing the two data types requires a co-located ionosonde and GNSS scintillation receiver. The study rests on two such stations, and their data gaps fall at different times (Fig.~\ref{fig:events_timeline}). These uneven gaps do not allow a fair comparison between the stations, so results are reported for both stations together. The record covers mostly the ascending and maximum phases of solar cycle 25, leaving low solar activity untested. Only 6 of the 409 event nights have their first onset after midnight LST, too few to assess post-midnight onsets separately. Thus, we report the results for post-sunset onsets only, and the gain of the ionogram for post-midnight onsets remains unresolved.

The station-level target is a deliberate simplification, chosen so that the gain reflects anticipation of the onset. The models forecast one $S_4$ series per station, without spatial structure, and a bubble counts as an event as soon as it reaches any LoS. They cannot track it beforehand and must anticipate the onset. A spatial model could also forecast scintillation on a LoS by following an existing bubble as it drifts toward that LoS depending on the lead time. Such tracking would make the GNSS baseline stronger, and the relative gain of the ionogram might be smaller. However for bubbles that form over the station, the ionogram would keep its advantage before the onset, as highlighted in this study.

Since the models here are not spatial, how far the gain will reach is another question. Our previous work describes a multi-instrument heterogeneous dynamic graph model as a feasible extension to dynamic graph modeling of GNSS observations~\cite{tur26}. Such a model with both ionogram and GNSS inputs would allow experiments on the spatial scale of the ionogram gain. It would also let us test whether each instrument covers its own cases: the ionosonde for bubbles that form nearby, and the GNSS network for those that drift in.

The $h'F$ experiment tests how much of the precursor information in the ionogram a scaled parameter carries. Adding $h'F$, $\dot{h}'F$ and $\ddot{h}'F$ to the Scalar model improves its performance over the onset to peak phase, but the gain does not match using the full image. Here we note that previous studies use different targets. Abadi et al.~\cite{Abadi22} make a day-to-day forecast, one answer per night on whether ESF or an EPB will occur, and Patra and Das~\cite{Patra25} similarly make one prediction per night, restricted to whether an EPB forms overhead before $F$-region sunset (excluding drifted bubbles that arrive later). Here, the models forecast at every 5-min step when scintillation reaches the station, while not differentiating bubbles that drift in. Results in Fig.~\ref{fig:when} suggest that precursors that separate event nights are not necessarily informative about the onset time of an upcoming scintillation related to plasma bubbles.

We attribute this difference to the image keeping relevant information beyond a single scaled parameter. Averaged over many events, the positive Shapley contributions before the onset lie at the low-frequency end of the $F$ trace (Fig.~\ref{fig:shap_frames}). Within this region, the model may also use the shape of the trace and early range spreading along the lower $F$ trace, which $h'F$ does not capture.

Lastly, these results also explain why the onset to peak gain is significant only in BSS (Table~\ref{tab:phase_results}). The ionogram raises the event probability before the onset. BSS scores the probability values, so it rewards this earlier rise. PR-AUC and ROC-AUC depend only on the ranking, and once $S_4$ starts to rise in the input window, the Scalar model ranks the event samples about as well as the Fusion model. This is partly due to PR-AUC and ROC-AUC calculated for all decision thresholds. Early rise we observe in the ionogram does not equal gains in every decision threshold. In practice, it is expected to have a better final forecast performance for the onset to peak phase provided the model uses a tuned decision threshold. Onset being the most important phase for operational users, the ionogram therefore adds value to early warning of scintillation events.

\section{Conclusions}
\label{sec:conclusions}

This study investigates whether ionogram images improve short-term forecasts of post-sunset scintillation at two SEALION stations near the magnetic equator, Chumphon in Thailand and Bac Lieu in Vietnam. On 230 test nights from both stations, the ionogram helps in the onset to peak phase of scintillation events. There, the Image and Fusion models significantly improve the BSS over the Scalar model by 26\% and 23\%, from 0.41 to 0.52 and 0.51. The gain grows with the lead time. From the 15- to the 120-min lead, the BSS of the Scalar model falls by 34\% and that of the Image model by 15\%. At 120~min, the Image and Fusion models exceed the Scalar model by 40\% and 32\%. The Image model at 120~min is about as skillful as the Scalar model at 30~min.

The gain is concentrated in forecasts issued in the hour before the onset. In this hour, the ionogram raises the event probability, mainly through frames recorded in the PRE window and through the low-frequency end of the F trace. At the 120-min lead, the Scalar model has no skill until the event appears in its GNSS input. The ionogram models have skill from about 40~min before the onset, and they lower the probability error of these forecasts by 28 to 36\% relative to the Scalar model. Adding $h'F$ and $\dot{h}'F$ to the Scalar model narrows this gap but does not close it. Against this variant, the ionogram models still lower the Brier score by 19 to 29\%. The ionogram image therefore carries scintillation onset information that neither the GNSS observables nor $h'F$ and its uplift rate provide.

These results come from two stations during the ascending and maximum phases of solar cycle 25 and do not cover post-midnight onsets. Within this scope, the ionosonde anticipates the onset and the GNSS receivers follow the event once it has started. How far from the station the ionogram informs the forecast, and its use in an operational forecasting model, are left for future work.

\section*{Acknowledgements}
This work was supported by Nanyang Technological University (NTU) and Thales Alenia Space. The authors wish to thank Erick Lansard, Yung-Sze Gan, Mathias Vanden Bossche, Sebastien Trilles, and Louis Hecker of Thales Alenia Space for their helpful discussions and support during the course of this work. The CPN ionosonde and GNSS receiver are operated by the National Institute of Information and Communications Technology (NICT), Japan, in collaboration with King Mongkut's Institute of Technology Ladkrabang (KMITL), Thailand. The BCL ionosonde and GNSS receiver are operated by NICT in collaboration with the Institute of Earth Sciences, Vietnam Academy of Science and Technology (IES-VAST), Vietnam. Claude (Anthropic) was used for programming, language and figure editing support during manuscript preparation. All scientific content, results, and conclusions are the authors' own.

\section*{Data Availability Statement}
SEALION ionograms, parameters and GNSS scintillation data are available from NICT at \url{https://aer-nc-web.nict.go.jp/sealion/} subject to the SEALION data policy. Multi-GNSS orbit products are available from Wuhan
University via the IGS MGEX data centers. OMNI data are available at \url{https://omniweb.gsfc.nasa.gov/}.
The algorithm developed for denoising ionograms is available at \url{https://github.com/Mert-chan/ionogram-denoiser}.

\bibliographystyle{IEEEtran}
\bibliography{ref}

\vfill

\end{document}